\documentclass[12pt]{article}
\usepackage{epsfig}

\def\fun#1#2{\lower3.6pt\vbox{\baselineskip0pt\lineskip.9pt
\ialign{$\mathsurround=0pt#1\hfil##\hfil$\crcr#2\crcr\sim\crcr}}}

\newcommand{\beq}{\begin{equation}}
\newcommand{\eeq}{\end{equation}}

\begin{document}

\title{Once again about the reaction $\phi (1020)\to \gamma \pi\pi$ }
\author{ V.V. Anisovich}
\date{June 26, 2006}
\maketitle

\begin{abstract}
Gauge invariance constraints and analytical properties of
the amplitudes
$e^+e^- \to \gamma\pi\pi$ and
$\phi (1020)\to \gamma\pi\pi$  are discussed in line
with the criticism of N.N. Achasov in hep-ph/0606003.  I show
that
the formulae  of our paper
(Yad. Fiz. {\bf 68} 1614 (2005) [PAN {\bf 68} 1554 (2005)],
hep-ph/0403123)  have nothing common
with those N.N. Achasov claims to be ours - his criticism is therefore
misplaced.

\end{abstract}

PACS numbers: 12.39.-x , 13.40.Hq , 13.65.+i

In the paper \cite{1}, we investigated the reactions
$e^+e^- \to \gamma\pi\pi$,
$\phi (1020)\to \gamma\pi\pi$
and came
to the conclusion that available experimental data  do not contradict
the assumed $q\bar q$ structure of $f_0(980)$. This paper was recently
criticized by N.N. Achasov in \cite{acha}. He wrote that "the
realization of gauge invariance  as a concequent of cancellation
between the $\phi (1020)\to \gamma f_0(980)\to \gamma\pi^0\pi^0$
resonance contribution and background one, suggested in Ref. [1], is
misleading."

In this short note, I would like to remind  the logic of reasoning in
Ref. \cite{1} and, comparing it with that of N.N. Achasov \cite{acha},
to argue in which points the logic  of N.N. Achasov is
incorrect.

\section{The logic of paper [1]}

Let us start with the general formula for the transition amplitude
$e^+e^-  \to \gamma\pi\pi$ assuming that the $e^+e^-$ system is in the
$1^{--}(V)$ state, $\pi\pi$ system in the $I=0,0^{++}(S)$ state and
the photon is real (eq. (20) in \cite{1}):
\beq
A_{\mu\alpha}^{(V\to\gamma S)}(s_V,s_S,q^2= 0)\, = \,
\left(g_{\mu\alpha}-\frac{2q_\mu P_{V\alpha}}{s_V-s_S}\right)
A_{V\to\gamma S}(s_V,s_S,0)\ .
\label{13}
\eeq
The indices $\mu$ and $\alpha$ refer to the initial vector state (total
momentum $P_V$ and $P_V^2=s_V$) and photon (momentum $q$ and $q^2= 0$).
We have $(P_V-q)^2=s_S$ and $(P_Vq)=(s_V-s_S)/2$. The spin operator
$[g_{\mu\alpha}-2q_\mu P_{V\alpha}/(s_V-s_S)]$  is gauge invariant:
$$
[g_{\mu\alpha}-2q_\mu P_{V\alpha}/(s_V-s_S)]q_\alpha=0, \quad
P_{V\mu}[g_{\mu\alpha}-2q_\mu P_{V\alpha}/(s_V-s_S)]=0.
$$
The requirement of analyticity (absence of the pole at $s_V=s_S$) leads
to the condition (eq. (21) in \cite{1}):
\beq
\bigg[A_{V\to\gamma S}(s_V,s_S,0)\bigg]_{s_V\to s_S}\ \sim \
 (s_V - s_S)\,  ,
\label{14}
\eeq
that is the threshold theorem for the transition amplitude
$V\to\gamma S$.

It should be now emphasized that the form of the spin operator in
eq.(\ref{13}),
 $[g_{\mu\alpha}-2q_\mu P_{V\alpha}/(s_V-s_S)]$,
 is not unique. Alternatively, one can write the spin factor
as a metric tensor
$g^{\perp\perp}_{\mu\alpha}$ which works in the space orthogonal to
$P_V$ and $q$, i.e.
$P_{V\mu}g^{\perp\perp}_{\mu\alpha}=0$ and
$g^{\perp\perp}_{\mu\alpha}\,q_\alpha=0$. Ambiguities in choice
of the spin operator for the process $V\to \gamma S$ are due to the
fact that the difference
$$
g^{\perp\perp}_{\mu\alpha}-
[g_{\mu\alpha}-2q_\mu P_{V\alpha}/(s_V-s_S)]
$$
is the nilpotent operator, for the detail see
\cite{maxim,operators} and eqs. (22)-(25) in \cite{1}. (Note that the
use of the operator $g^{\perp\perp}_{\mu\alpha}$ was also criticized by
N.N. Achasov \cite{acha02} but he did not pay attention to the presence
of nilpotent operators).

\begin{figure}[t]
\centerline{\epsfig{file=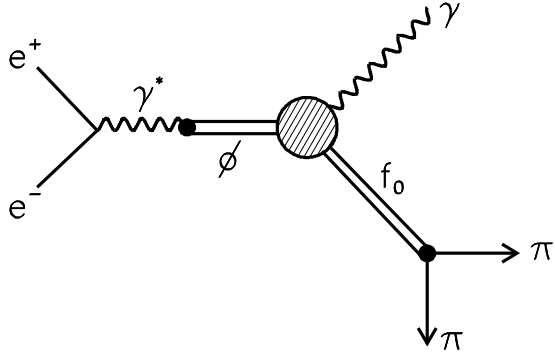,width=8cm}}
\caption{Process $e^+e^-\to \gamma\pi\pi$: residues in the
$e^+e^-$ and $\pi\pi$ channels determine the
$\phi \to \gamma f_0$ amplitude.}
\end{figure}

\subsection {Transitions $e^+e^-\to  \gamma\pi\pi$ ,
 $ \phi \to \gamma f_0$ and $ \phi \to \gamma\pi\pi $
.}

The amplitude of the transition $e^+e^-\to  \gamma\pi\pi$ determines
the amplitudes
 $ \phi \to \gamma f_0$ and $ \phi \to \gamma\pi\pi $ as
 corresponding residues of the pole terms.

\subsubsection {The amplitude $e^+e^-\to  \gamma\pi\pi$: the pole and
 background terms}

The  $A_{\phi\to\gamma f_0}$
amplitude is defined as
the amplitude  residue of the corresponding double-pole term in the
 amplitude $e^+e^-\to  \gamma\pi\pi$.
For
$e^+e^-\to\gamma\pi\pi$, see Fig. 1, the amplitude
with singled out double-pole term reads (eq. (26) in \cite{1}):
\begin{eqnarray}
&& A^{(e^+e^-\to\gamma\pi\pi)}_{\mu\alpha} (s_V,s_S,0)\ =\left(
g_{\mu\alpha}-\frac{2q_\mu P_{V\alpha}}{s_V-s_S}\right)\times
\nonumber \\
&&
\times
 \left[G_{e^+e^-\to\phi}\
\frac{A_{\phi\to\gamma f_0}(m^2_\phi,m^2_{f_0},0)}{(s_V-m^2_\phi)
(s_S-m^2_{f_0})}\ g_{f_0\to\pi\pi} +B(s_V,s_S,0)\right].
\label{19}
\end{eqnarray}
Here,
$A(m^2_\phi,m^2_{f_0},0)$, up to the factors $G_{e^+e^-\to\phi}$ and
$g_{f_0\to\pi\pi}$, is the residue in the amplitude poles
 $s_V=m^2_\phi$ and
$s_S=m^2_{f_0}$: just this value supplies us with the transition
amplitude for  the reaction with
bound states, $\phi\to\gamma f_0$.

In eq. (\ref{19}),
we deal with the double-pole term only. The amplitude
$B(s_V,s_S,0)$  may contain  single-pole terms too. Writing down
all pole terms, one has:
\begin{eqnarray} &&
A^{(e^+e^-\to\gamma\pi\pi)}_{\mu\alpha} (s_V,s_S,0)\ =\left(
g_{\mu\alpha}-\frac{2q_\mu P_{V\alpha}}{s_V-s_S}\right)
 \left[G_{e^+e^-\to\phi}\
\frac{A_{\phi\to\gamma f_0}(m^2_\phi,m^2_{f_0},0)}{(s_V-m^2_\phi)
(s_S-m^2_{f_0})}\ g_{f_0\to\pi\pi} + \right.
\nonumber\\
&&\left. + G_{e^+e^-\to\phi}\
\frac{B_{\phi}(m^2_\phi,s_S,0)}{s_V-m^2_\phi}+
\frac{B_{f_0} (s_V,m^2_{f_0},0)}{s_S-m^2_{f_0}}\ g_{f_0\to\pi\pi}
+B_0(s_V,s_S,0)\right].
\label{19a}
\end{eqnarray}
The threshold theorem (\ref{14}) for the amplitude (\ref{19a}) reads:
\begin{eqnarray} &&
 \left[G_{e^+e^-\to\phi}\
\frac{A_{\phi\to\gamma f_0}(m^2_\phi,m^2_{f_0},0)}{(s_V-m^2_\phi)
(s_V-m^2_{f_0})}\ g_{f_0\to\pi\pi} + \right.
\nonumber\\
&&\left. + G_{e^+e^-\to\phi}\
\frac{B_{\phi}(m^2_\phi,s_V,0)}{s_V-m^2_\phi}+
\frac{B_{f_0} (s_V,m^2_{f_0},0)}{s_V-m^2_{f_0}}\ g_{f_0\to\pi\pi}
+B_0(s_V,s_V,0)\right]=0\, .
\label{19a-t}
\end{eqnarray}

\subsubsection {The $\phi\to  \gamma f_0$ amplitude }

 If we deal with stable
composite particles, in other words, if $\phi$ and $f_0$ can be
included into the set of fields $|in\rangle$ and $\langle out|$, the
transition amplitude $\phi\to\gamma f_0$ can be written in the form
similar to  (\ref{13}) (see eq. (27) in \cite{1}):
\beq
\label{20}
A^{(\phi\to\gamma f_0)}_{\mu\alpha}(m^2_\phi,m^2_{f_0},0)\
=\left(g_{\mu\alpha} -\frac{2q_\mu
P_{\phi\alpha}}{m^2_\phi-m^2_{f_0}}\right)
A_{\phi\to\gamma f_0}\left(m^2_\phi,m^2_{f_0},0\right).
\eeq
Here, we have substituted
$P_V\to P_{\phi}$ where
$$
P_{\phi}^2=m^2_\phi, \quad
(P_{\phi}-q)^2=m^2_{f_0}, \quad q^2=0.
$$
For $A_{\phi\to\gamma
f_0}\left(m^2_\phi,m^2_{f_0},0\right)$, the threshold theorem reads (eq.
(28) in \cite{1}):
\beq
\label{21}
\left[A_{\phi\to\gamma
f_0}(m^2_\phi,m^2_{f_0},0)\right]_{m^2_\phi \to m^2_{f_0}}\ \sim\
(m^2_\phi-m^2_{f_0})\ ,
\eeq
that means that the threshold theorem of
eq. (\ref{21}) reveals itself as a requirement of analyticity of the
amplitude (\ref{20}).

\subsubsection {The $\phi\to  \gamma \pi\pi$ amplitude}

To describe
the reaction $\phi\to\gamma \pi\pi$ considering $\phi$ as a stable
particle, one needs to separate from (\ref{19a}) the  pole factors:
$\sim (s_V-m^2_\phi)^{-1}$.
The amplitude $\phi\to\gamma \pi\pi$ is the residue in the
pole $s_V=m^2_\phi $:
\begin{eqnarray}
A^{(\phi\to\gamma\pi\pi)}_{\mu\alpha} (m^2_\phi,s_S,0)\ =\left(
g_{\mu\alpha}-\frac{2q_\mu P_{\phi\alpha}}{m^2_\phi-s_S}\right)
 \left[
\frac{A_{\phi\to\gamma f_0}(m^2_\phi,m^2_{f_0},0)}{
s_S-m^2_{f_0}}\ g_{f_0\to\pi\pi}
+
B_{\phi}(m^2_\phi,s_S,0)\right],
\label{19b}
\end{eqnarray}
where
$$
P_{\phi}^2=m^2_\phi, \quad
(P_{\phi}-q)^2=s_S, \quad q^2=0.
$$
Let us emphasize once more that
$B_{\phi}(m^2_\phi,s_S,0)$ does not contain pole terms and
$m_\phi$ is fixed ($m_\phi=1020$ MeV).

The analyticity condition reads:
\begin{eqnarray}
 \left[
\frac{A_{\phi\to\gamma f_0}(m^2_\phi,m^2_{f_0},0)}{
s_S-m^2_{f_0}}\ g_{f_0\to\pi\pi}
+
B_{\phi}(m^2_\phi,s_S,0)\right]_{s_S\to m^2_\phi}\sim (s_S - m^2_\phi) .
\label{19c}
\end{eqnarray}

\subsection{ Example of idealistic description of
$\phi (1020)\to\gamma\pi\pi $}

To make clear our way of calculations, let us consider
the idealistic case: the $f_0$ is a standard Breit-Wigner resonance,
while $K\bar K$ channel is stongly supressed in the region under
consideration and may be neglected. The $\phi$ is considered as a
stable particle (the width of $\phi (1020)$ is small).

In this case, the $\phi\to\gamma \pi\pi$ amplitude  is given
by eq. (\ref{19b}), with
\beq
\label{1-1}
 m_{f_0}^2= M_{0}^2-i\Gamma M_{0} \ .
\eeq
To be illustrative, let us rewrite (\ref{19b}) extracting the
Breit-Wigner resonance pole explicitly:
 \begin{eqnarray}
 A^{(\phi\to\gamma\pi\pi)}_{\mu\alpha}
(m^2_\phi,s_S,0)\ =\left( g_{\mu\alpha}-\frac{2q_\mu
P_{\phi\alpha}}{m^2_\phi-s_S}\right)
 \left[
\frac{A_{\phi\to\gamma f_0}(m^2_\phi, m_{f_0}^2,0)}{
s_S-M_{0}^2+i\Gamma M_{0}}\ g_{f_0\to\pi\pi}
+
B_{\phi}(m^2_\phi,s_S,0)\right],
\label{1-1a}
\end{eqnarray}
Recall that $B_{\phi}(m^2_\phi,s_S,0)$ is the background contribution,
i.e. it does not contain the pole term related to $f_0(980)$ and
$m_\phi$ is fixed, $m_\phi=1020$ MeV. In the fitting procedure,
$B_{\phi}(m^2_\phi,s_S,0)$ can be approximated by a smooth term or as a
contribution of other poles (for example, such as $\sigma$). The
fitting to the
parameters of $B_{\phi}(m^2_\phi,s_S,0)$ should be carried out under
two constraints:
\begin{eqnarray}
 &&  \left (
\frac{A_{\phi\to\gamma f_0}(m^2_\phi,m^2_{f_0},0)}{
s_S-M_{0}^2+i\Gamma M_{0}}\ g_{f_0\to\pi\pi}
+
B_{\phi}(m^2_\phi,s_S,0)\right)=
\nonumber \\
&&
= \left |
\frac{A_{\phi\to\gamma f_0}(m^2_\phi,m^2_{f_0},0)}{
s_S-M_{0}^2+i\Gamma M_{0}}\ g_{f_0\to\pi\pi}
+
B_{\phi}(m^2_\phi,s_S,0)\right | \exp{\left (i\delta_0^0(s_S)\right )}
\label{1-2}
\end{eqnarray}
and
\begin{eqnarray}
\frac{A_{\phi\to\gamma f_0}(m^2_\phi,m^2_{f_0},0)}{
m^2_\phi-M_{0}^2+i\Gamma M_{0}}\ g_{f_0\to\pi\pi}
+
B_{\phi}(m^2_\phi,m^2_\phi,0)=0 \ .
\label{1-3}
\end{eqnarray}
The factor $\exp{\left (i\delta_0^0(s_S)\right )}$,
where $\delta_0^0(s_S)$ is the $\pi\pi$ scattering phase shift, appears
in (\ref{1-2})
because of the final state interactions of pions, while the condition
(\ref{1-3}) results from the threshold theorem (\ref{19c}).

\subsection{ Description of the reaction
$\phi (1020)\to\gamma\pi\pi $ in paper [1]}

The vector meson $\phi(1020)$ has rather small decay width,
$\Gamma_{\phi(1020)} \simeq4.5\,$MeV; from this point of view there is
no doubt that treating $\phi(1020)$ as a stable particle
is  reasonable. As to $f_0(980)$, the picture is not so determinate. In
the PDG compilation
\cite{PDG}, the $f_0(980)$ width is given in the interval
$40\le\Gamma_{f_0(980)}\le100\,$MeV, and the width uncertainty
is related not to the data unaccuracy (experimental data are
rather good) but is due to a vague definition of the width.

\begin{figure}[h]
\centerline{\epsfig{file=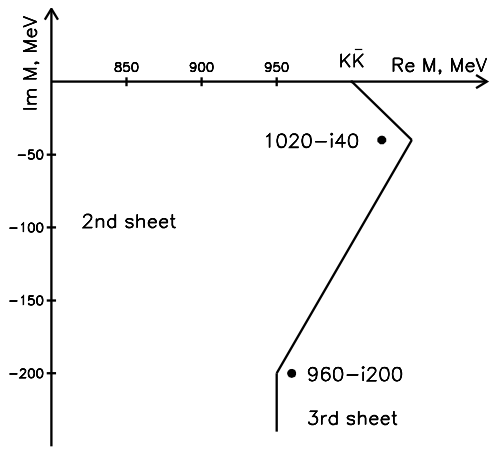,width=7cm}}
\caption{Complex-$M$ plane and location of the poles corresponding to
$f_0(980)$; the cut related to the $K\bar K$ threshold
is shown as a broken solid line (here we denote $M=\sqrt{s_S}$).}
\end{figure}

The definition
of the $f_0(980)$ width is aggravated by the $K\bar K$
threshold singularity that leads to the existence of two, not one,
poles. According to the $K$-matrix analyses \cite{kmat,ufn}, there are
two poles in the  $(IJ^{PC}=00^{++})$-wave at $s\sim 1.0$ GeV$^2$,
\beq
M_I \simeq
1.020-i0.040\mbox{ GeV }, \quad
M_{II} \simeq0.960-i0.200\mbox{ GeV },
\label{35}
\eeq
which are
located on different complex-$M$ sheets related to the
$K\bar K$-threshold, see Fig. 2.

A significant trait of the $K$-matrix analysis is that it also gives
us, along with the characteristics of real resonances, the positions of
levels before the onset of the decay channels, i.e. it determines the
bare states. In addition, the $K$-matrix analysis allows us to observe
the transform of bare states into real resonances. In Fig. 3,
one can see
such a transform of the $00^{++}$-amplitude poles
by switching off the decays
$f_0\to\pi\pi,K\bar K,\eta\eta,\eta\eta',\pi\pi\pi\pi$.
One may see that, after switching off the decay channels, the
$f_0(980)$ turns into stable state, approximately  300 MeV lower:
\beq
f_0(980)\ \longrightarrow\ f^{bare}_0(700\pm100)\ .
\label{36}
\eeq

\begin{figure}[h]
\centerline{\epsfig{file=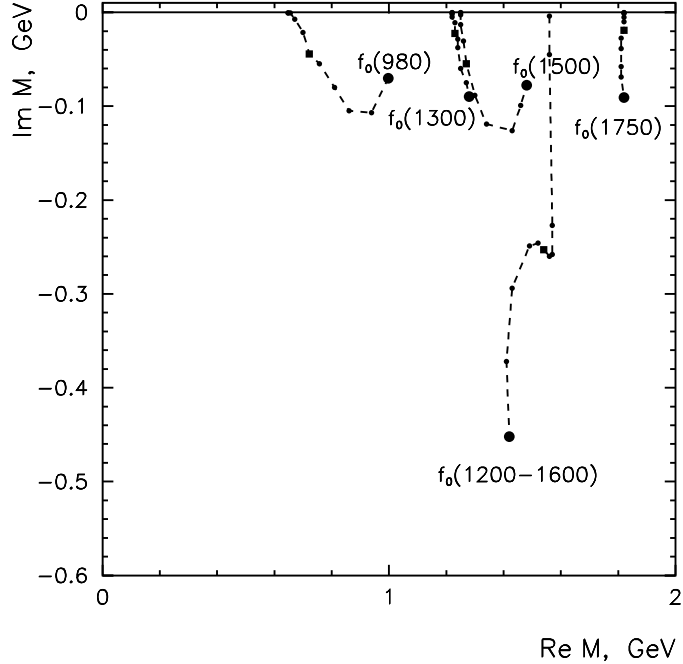,width=10cm}}
\caption{Complex-$M$ plane: trajectories of poles
corresponding to the states $f_0(980)$, $f_0(1300)$, $f_0(1500)$,
$f_0(1750)$, $f_0(1200-1600)$ within a uniform onset of the decay
channels.}
\end{figure}

\begin{figure}[h]
\centerline{\epsfig{file=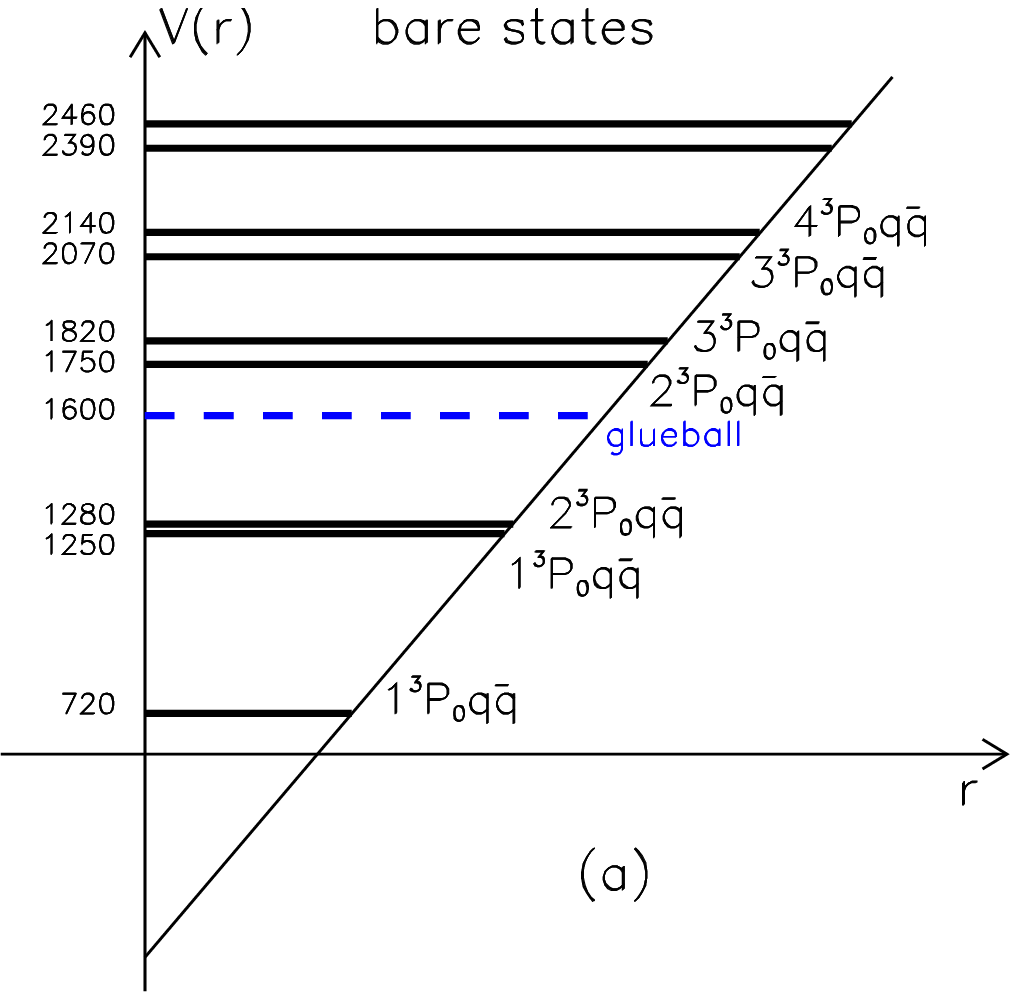,width=9cm}
            \epsfig{file=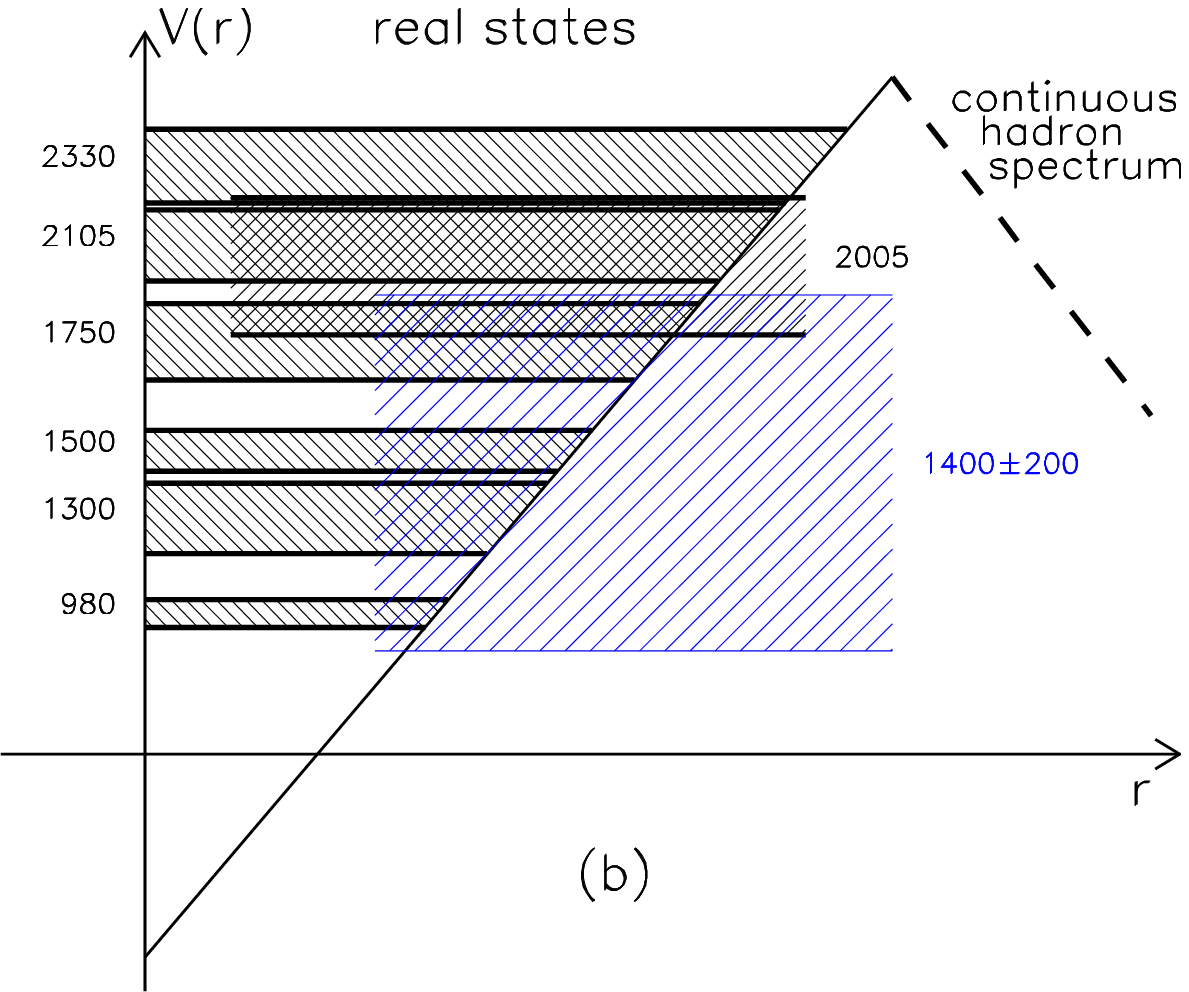,width=9cm}}
\caption{The $f_0$-levels in the potential well depending on
the onset of the decay channels: bare states (a)  and real resonances
(b).}
\end{figure}

The transform of bare states into real resonances may be illustrated
by Fig. 4 for the levels in the potential well: bare states are the
levels
in a well with impenetrable wall (Fig. 4a); at the onset of the decay
channels (under-barrier transitions, Fig. 4b) the stable levels
transform into real resonances. Note, that in this process
one resonance (in the case of the $00^{++}$ states, it is gluonium)
accumulates the widths of the neighbouring resonances thus becoming
the broad state (the effect of accumulation of widths was firstly
seen in nucler physics \cite{accum}).

The $K$-matrix amplitude of the  $00^{++}$-wave reconstructed in
\cite{kmat} gives us the possibility to trace the evolution of the
transition form factor
$\phi(1020)\to\gamma f^{bare}_0(700\pm100)$ during the transformation of
the bare state  $f^{bare}_0(700\pm100)$ into the $f_0(980)$ resonance.
Using the diagrammatic language, one can say that the
evolution of the form factor
$F^{(bare)}_{\phi\to\gamma f_0}$ is due to the processes shown in Fig.
5: $\phi$-meson goes into $f^{bare}_0(n)$, with the emission of the
photon, then $f^{bare}_0(n)$ decays into mesons
$f^{bare}_0(n)\to h_ah_a=\pi\pi$,
$K\bar K$, $\eta\eta$, $\eta\eta'$, $\pi\pi\pi\pi$. The decay
yields may rescatter thus coming to  final states.

\begin{figure}[h]
\centerline{\epsfig{file=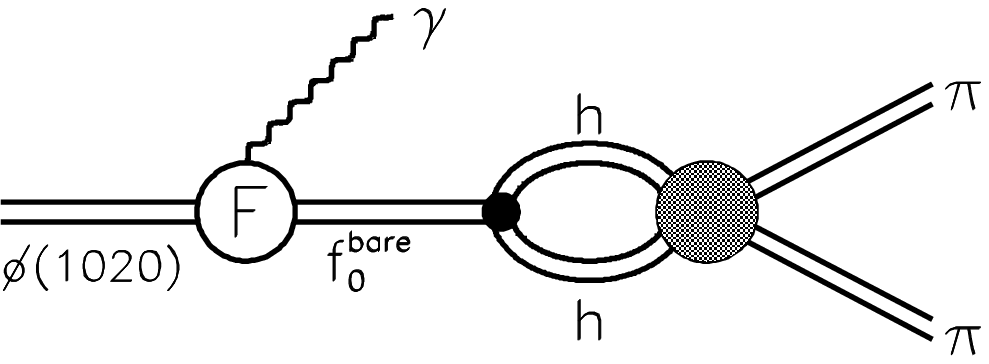,width=10cm}}
\caption{Diagram for the transition $\phi (1020)\to \gamma \pi\pi$
with final state interaction taken in terms of  the K-matrix
representation (the right-hand side block $hh\to \pi\pi$).}
\end{figure}

With the use of the K-matrix technique, the amplitude
$\phi(1020)\to\gamma\pi\pi$ is given by eq. (\ref{19b}) with the
following
replacement (see Fig. 5):
\begin{eqnarray}
\label{37}
&& \left[
\frac{A_{\phi\to\gamma f_0}(m^2_\phi,m^2_{f_0},0)}{
s_S-m^2_{f_0}}\ g_{f_0\to\pi\pi}
+
B_{\phi}(m^2_\phi,s_S,0)\right] \longrightarrow \\
&&\longrightarrow
\sum_{a} \left( \sum_{n}
\frac{F^{(bare)}_{\phi (1020)
\to\gamma f_0^{bare}(n)}\,g^{bare}_a(n)}{M^2_n-s_S}+b_a(s_S)\right)
\left(\frac1{1-i\hat\rho(s_S)\hat K(s_S)}\right)_{a,\pi\pi}\ ,
\nonumber
\end{eqnarray}
where the  $K$-matrix
elements $K_{ab}(s_S)$ contain the poles corresponding to bare
states:
\beq
K_{ab}(s)=\sum_n \frac{g^{bare}_a(n)\,
g^{bare}_b(n)}{M^2_n-s_S}+f_{ab}(s_S).
\label{37a}
\eeq
Here $M_n$ is the mass of bare state, $g^{bare}_a(n)$ is the coupling
for the transition
$f^{bare}_0(n)\to a$, where
 \beq
 a=\pi\pi,\ K\bar K,\ \eta\eta,\
\eta\eta',\ \pi\pi\pi\pi .
\label{37a-1}
\eeq
The matrix element
$(1-i\hat\rho(s)\hat K(s_S))^{-1}$ takes into account of the
rescattering of the formed mesons. Here $\hat\rho(s_S)$ is the diagonal
matrix of phase spaces for hadronic states (for example, for the
$\pi\pi$ system it reads:
$\rho_{\pi\pi}(s_S)=\sqrt{(s_S-4m^2_\pi)/s_S}$ ).
The functions $b_a(s_S)$ and $f_{ab}(s)$ describe background
contributions, they are smooth ones in the right-hand side
half-plane, at Re$\,s_S>0$.

The formulae (\ref{37}) and (\ref{37a})  are presented in \cite{1} (eqs.
(46),(47)), with the renotation $s_S\to s$.

To be scrupulous, let us present the amplitude
$\phi(1020)\to\gamma\pi\pi$ explicitly:
\begin{eqnarray} &&
A^{(\phi\to\gamma\pi\pi)}_{\mu\alpha} (m^2_{\phi},s_S,0)\ =\left(
g_{\mu\alpha}-\frac{2q_\mu P_{\phi\alpha}}{m^2_{\phi}-s_S}\right)\times
\nonumber \\
&&\times
\left[\sum_{a} \left( \sum_{n}
\frac{F^{(bare)}_{\phi (1020)
\to\gamma f_0^{bare}(n)}\, g^{bare}_a(n)}{M^2_n-s_S}+b_a(s_S)\right)
\left(\frac1{1-i\hat\rho(s_S)\hat K(s_S)}\right)_{a,\pi\pi}\right]\
. \label{37b}
\end{eqnarray}
Therefore, the threshold condition reads:
\beq \label{37c}
\left[\sum_{a} \left( \sum_{n}
\frac{F^{(bare)}_{\phi (1020)
\to\gamma f_0^{bare}(n)}\, g^{bare}_a(n)}{M^2_n-s_S}+b_a(s_S)\right)
\left(\frac1{1-i\hat\rho(s_S)\hat K(s_S)}\right)_{a,\pi\pi}
\right]_{s_S \to m^2_\phi}   \sim m^2_{\phi}- s_S \ .
\eeq
The fitting procedure of the reaction $\phi\to\gamma\pi\pi$
should be performed with the threshold constraint only:
\beq \label{37c1}
\left[\sum_{a} \left( \sum_{n}
\frac{F^{(bare)}_{\phi (1020)
\to\gamma f_0^{bare}(n)}}{M^2_n- m^2_{\phi}}\,g^{bare}_a(n)+b_a( m^2_{\phi})\right)
\left(\frac1{1-i\hat\rho( m^2_{\phi})\hat K( m^2_{\phi})}\right)_{a,\pi\pi}
\right]=0
\eeq
because here the final state interaction is taken into account
explicitly (in contrast to eq. (\ref{19b}) where one needs to account
for the constraint (\ref{1-2})).

Formula (\ref{37b}) was used in \cite{1} for the calculation of
residues in the poles $s_S=M^2_I$ and $s_S=M^2_{II}$, see (\ref{35})
and Fig. 2. The residues are $A_{\phi\to\gamma
f_0}(m^2_\phi,s_S=M^2_{I},0)$  and $A_{\phi\to\gamma
f_0}(m^2_\phi,s_S=M^2_{II},0)$ (see Section 5.3 in \cite{1}) which
characterize the production of the considered state; this calculation
was in line with singling out the pole in (\ref{19b}). (Discussion
of the role of the pole residues can be found, for example, in
\cite{ufn,penn} and references therein). Then, we compare the amplitude
$A_{\phi\to\gamma f_0}(m^2_\phi,s_S=M^2_I,0)$ (for the pole which is
the nearest one to the physical region) with quark model predictions
(Section 7) and conclude that experimental data on the reaction
$\phi (1020)\to\gamma f_0(980)$ do not contradict the suggestion about
the dominance of the $q\bar q$ component in $f_0(980)$.

\subsubsection{Illustrative examples of the K-matrix description of
        the decay  $\phi \to\gamma \pi\pi $ }

Following \cite {kmat}, the K-matrix consideration of the decay
$\phi (1020)\to\gamma \pi\pi$ was performed in \cite{1} with the use of
five channels (\ref{37a-1}) and five resonance states in the
$00^{++}$ wave.

To make the reader more acquainted with this method,  consider
illustative examples similar to that given in Section 1.1. We present
formulae for the two cases when in the $00^{++}\pi\pi$ channel we have
(i) one resonance and (ii) two of them.

{\bf (i) One resonance in the $00^{++}\pi\pi$ channel. }

The decay amplitude $\phi \to\gamma \pi\pi $ in the K-matrix
representation is written as
\begin{eqnarray} &&
A^{(\phi\to\gamma\pi\pi)}_{\mu\alpha} (m^2_{\phi},s_S,0)\ =\left(
g_{\mu\alpha}-\frac{2q_\mu P_{\phi\alpha}}{m^2_{\phi}-s_S}\right)\times
\nonumber \\
&&\times
\left(
\frac{F^{(bare)}_{\phi \to\gamma f_0^{bare}}\,
g^{(bare)}_{\pi\pi}}{M^2_0-s_S}+b(s_S)\right)
\left(
1-i\rho_{\pi\pi}(s_S)[\frac{g^{(bare)\, 2}_{\pi\pi}}{M^2_0-s_S}+
f(s_S)]\right)^{-1} \ .
\label{37-1}
\end{eqnarray}
The first factor in the right-hand side of (\ref{37-1}) describes a
direct production of $\pi\pi$, while the second one is due to
rescattering of pions with the K-marix factor equal to
$$
K_{\pi\pi\to\pi\pi}=
\frac{g^{(bare)\, 2}_{\pi\pi}}{M^2_0-s_S}+
f(s_S) .
$$
Note that the form factor
$F^{(bare)}_{\phi \to\gamma f_0^{bare}}$ and coupling
$g^{(bare)}_{\pi\pi}$ do not depend on $s_S$, they are constant.
The functions $b(s_S)$ and $f(s_S)$ are smooth ones in the region under
consideration.

The threshold theorem for (\ref{37-1}) reads:
\beq
\left(
\frac{F^{(bare)}_{\phi \to\gamma f_0^{bare}}\,
g^{(bare)}_{\pi\pi}}{M^2_0-m^2_\phi}+b(m^2_\phi)\right)
\left(
1-i\rho_{\pi\pi}(m^2_\phi)[\frac{g^{(bare)\, 2}_{\pi\pi}}{M^2_0-m^2_\phi}+
f(m^2_\phi)]\right)^{-1}=0 \ .
\label{37-2}
\eeq

One may rewrite (\ref{37-1}) in the form similar to that of the
Breit-Wigner resonance:
\begin{eqnarray} &&
A^{(\phi\to\gamma\pi\pi)}_{\mu\alpha} (m^2_{\phi},s_S,0)\ =\left(
g_{\mu\alpha}-\frac{2q_\mu P_{\phi\alpha}}{m^2_{\phi}-s_S}\right)\times
\nonumber \\
&&\times
\frac{F^{(bare)}_{\phi \to\gamma f_0^{bare}}\,
g^{(bare)}_{\pi\pi}+b(s_S)(M^2_0-s_S)}
{M^2_0-s_S
-i\rho_{\pi\pi}(s_S)[g^{(bare)\, 2}_{\pi\pi}+
f(s_S)(M^2_0-s_S)]} \ .
\label{37-3}
\end{eqnarray}
The position of the $f_0$-resonance is determined by zero of the
denominator in (\ref{37-3}):
\beq
 M^2_0-s_S -i\rho_{\pi\pi}(s_S)[g^{(bare)\,
2}_{\pi\pi}+ f(s_S)(M^2_0-s_S)]=0 \ .
\label{37-4}
 \eeq
 Let the resonance
pole exist at
\beq
s_S=m^2_{Res}-i\Gamma_{Res}m_{Res} \equiv
M^2_{Res}\, .
\label{37-4-1}
\eeq
With this definition, one can rewrite
(\ref{37-3}) in the form of eq. (\ref{19b}):
\beq \label{37-5}
A^{(\phi\to\gamma\pi\pi)}_{\mu\alpha} (m^2_{\phi},s_S,0)\ =\left(
g_{\mu\alpha}-\frac{2q_\mu P_{\phi\alpha}}{m^2_{\phi}-s_S}\right)
\left[\frac{F^{(bare)}_{\phi \to\gamma f_0^{bare}}\,
g^{(bare)}_{\pi\pi}+b(M^2_{Res})}{M^2_{Res}-s_S}
+
B_{\phi}(m^2_\phi,s_S,0)\right]
\eeq
where
\beq
 \label{37-6}
B_{\phi}(m^2_\phi,s_S,0) =
\frac{F^{(bare)}_{\phi \to\gamma f_0^{bare}}\,
g^{(bare)}_{\pi\pi}+b(s_S)(M^2_0-s_S)}
{M^2_0-s_S
-i\rho_{\pi\pi}(s_S)[g^{(bare)\, 2}_{\pi\pi}+
f(s_S)(M^2_0-s_S)]}
-\frac{F^{(bare)}_{\phi \to\gamma f_0^{bare}}\,
g^{(bare)}_{\pi\pi}+b(M^2_{Res})}{M^2_{Res}-s_S} \, .
\eeq
The function $B_{\phi}(m^2_\phi,s_S,0)$ determined by (\ref{37-6})
does not contain the resonance pole. The theshold theorm (\ref{37-2})
reads now as a cancellation of the pole and the smooth background
term $B_{\phi}(m^2_\phi,s_S,0)$  at $s_S=m^2_\phi $.

{\bf (ii) Two resonances in the channel $00^{++}\pi\pi$. }

One may try to describe the $\pi\pi$ background by a broad resonance.
In the case of two resonances
the decay amplitude  in the K-matrix
representation for $\phi \to\gamma \pi\pi $ gives us:
\begin{eqnarray} &&
A^{(\phi\to\gamma\pi\pi)}_{\mu\alpha} (m^2_{\phi},s_S,0)\
=
\nonumber \\
&&=
\left(
g_{\mu\alpha}-\frac{2q_\mu P_{\phi\alpha}}{m^2_{\phi}-s_S}\right)
\left(
\frac{F^{(bare)}_{\phi \to\gamma f_0^{bare}(1)}\,
g^{(bare)}_{\pi\pi}(1)}{M^2_1-s_S}+
\frac{F^{(bare)}_{\phi \to\gamma f_0^{bare}(2)}\,
g^{(bare)}_{\pi\pi}(2)}{M^2_2-s_S}
+b(s_S)\right)
\times
\nonumber \\
&&\times
\left(
1-i\rho_{\pi\pi}(s_S)[\frac{g^{(bare)\, 2}_{\pi\pi}(1)}{M^2_1-s_S}+
\frac{g^{(bare)\, 2}_{\pi\pi}(2)}{M^2_2-s_S}+
f(s_S)]\right)^{-1} \ ,
\label{37-7}
\end{eqnarray}
with the following threshold condition:
\begin{eqnarray} &&
\left(
\frac{F^{(bare)}_{\phi \to\gamma f_0^{bare}(1)}\,
g^{(bare)}_{\pi\pi}(1)}{M^2_1-m^2_\phi}+
\frac{F^{(bare)}_{\phi \to\gamma f_0^{bare}(2)}\,
g^{(bare)}_{\pi\pi}(2)}{M^2_2-m^2_\phi}
+b(m^2_\phi)\right)
\times
\nonumber \\
&&\times
\left(
1-i\rho_{\pi\pi}(m^2_\phi)\left[\frac{g^{(bare)\,
2}_{\pi\pi}(1)}{M^2_1-m^2_\phi}+ \frac{g^{(bare)\,
2}_{\pi\pi}(2)}{M^2_2-m^2_\phi}+ f(m^2_\phi)\right]\right)^{-1}\, =\, 0
\ . \label{37-8}
\end{eqnarray}

\subsection{Approximate description of the $\pi^0\pi^0$ spectra
in $\phi\to\gamma\pi\pi$ with Flatt\'e formula for
$f_0(980)$}

The $\pi^0\pi^0$ spectrum in the reaction $\phi\to\gamma\pi^0\pi^0$
\cite{nov1} is shown in Fig. 6.

\begin{figure}[h]
\centerline{\epsfig{file=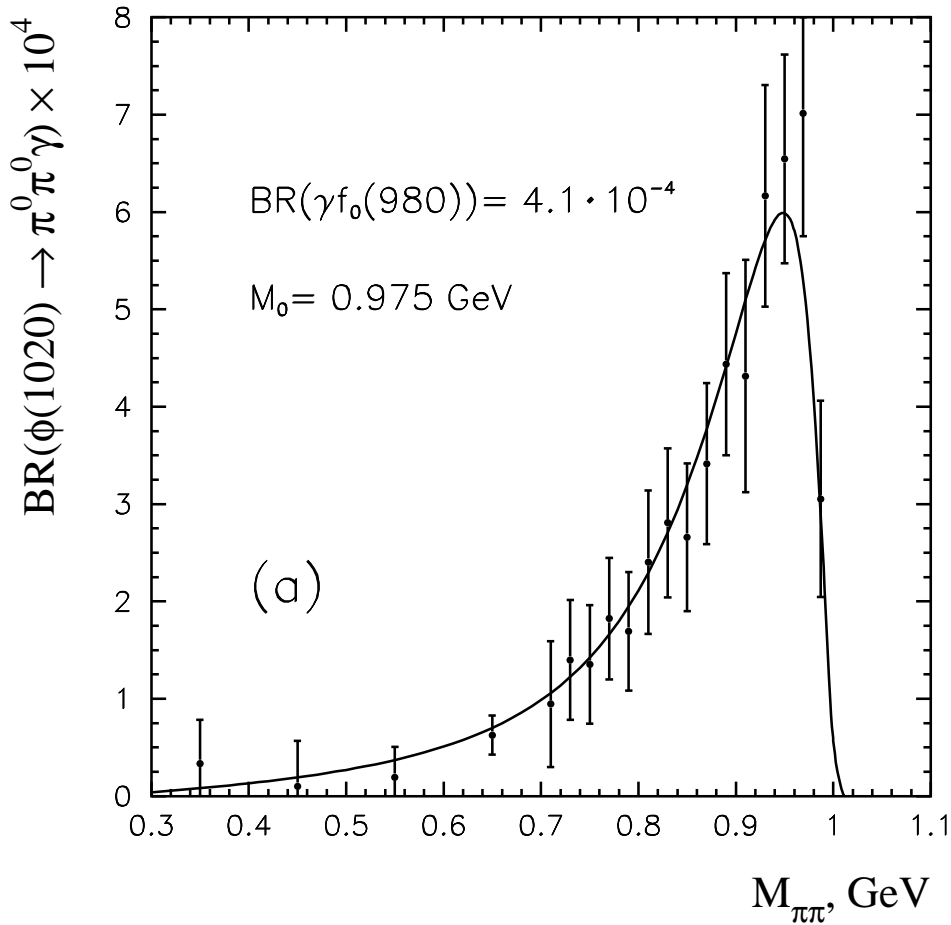,width=0.8\textwidth}}
\vspace{-0.5cm}
\caption{The $\pi\pi$ spectrum of the reaction $\phi(1020)\to
\gamma\pi^0\pi^0$ calculated with the Flatte\'e formula
(notation for the $\pi\pi$ invariant mass is redenoted here,
$\sqrt{s_S}\to M_{\pi\pi}$).
} \end{figure}

The resonance $f_0(980)$ has two dominant decay channels
$f_0(980)\to \pi\pi\,$, $K\bar K$, so the precise description of the
$\pi\pi$ spectrum needs  the K-matrix technique. But the
K-matrix description requires  more information, in particular, about
the reaction  $f_0(980)\to \gamma K\bar K$, that is not available now.
Therefore, a reasonable compromise may be the use of the
Breit--Wigner-type  formula, where the $K\bar K$ threshold
singularity is taken into account: it is the Flatt\'e formula
\cite{Flatte} or that suggested in \cite{content} (where the transition
length is taken into account).

In case of using  the Flatt\'e formula, the
reaction $\phi(1020)\to \gamma\pi^0\pi^0$ is described by
formulae (\ref{1-1a}) - (\ref{1-3}) of Section 1.2, with a change  of
the Breit-Wigner factor:
\beq
\label{sp5}
\frac{1}{s_S-M_{0}^2+i\Gamma M_{0}}\longrightarrow
\frac{1}{
s_S-M^2_0+ig_\pi^2
\rho_{\pi\pi}(s_S)
+ig_K^2\rho_{K\bar K}(s_S)},
\eeq
where $\rho_{K\bar K}(s_S)=\sqrt{(s_S-4m^2_K)/s_S}$ is the $K\bar K$
phase space.

In \cite{1}, the amplitude  $A_{\phi\to\gamma f_0}(m^2_\phi,
m_{f_0}^2,0)$ (see (\ref{1-1a})) was determined
 supposing the $q\bar q$ structure for $f_0(980)$.
Fitting to the $\pi\pi$ spectrum, see Fig. 6 \cite{nov1}, was performed
under the constraints (\ref{1-2}) and (\ref{1-3}), and the background
term in \cite{1} was parametrized as follows:
\beq \label{sp6}
B_{\phi}(m^2_\phi,s_S,0)
= C(s_S)\left [1+a(s_S - m^2_\phi)\right ]
\exp{\left [-\frac{ m^2_\phi -s_S}{\mu^2}\right ]}\ .
\eeq
The result is shown in Fig. 6.

Note that the condition (\ref{1-2}) is not valid in the region
$$
2m_K\leq \sqrt{s_S}\leq m_\phi\ ,
$$
but at such $\sqrt{s_S}$ the contribution from $\sigma(\phi\to
\pi^0\pi^0)$ is negligently small as compared to error bars in Fig. 6.

\section{The logic of paper \cite{acha}}

The aim of paper \cite{acha} is to demonstrate that formulae used in our
 paper \cite{1} are incorrect. To this aim, N.N. Achasov starts from
 our formula for $\phi \to \gamma \pi\pi$:
\beq
\label{acha1}
\sigma(\phi \to \pi\pi) \sim
\left| \frac{1}{
\label{acha3}s_S-M^2_0+ig_\pi^2
\rho_{\pi\pi}(s_S)
+ig_K^2\rho_{K\bar K}(s_S)}+B_{\phi}(m^2_\phi,s_S,0)\right|^2 .
\eeq
Then, he generalizes it for the reaction
$e^+e^-\to \phi\to\gamma\pi^0\pi^0$ in the following way (see eq. (9)
of \cite{acha}):
\begin{eqnarray}
&&
\sigma(e^+e^-\to \phi\to\gamma\pi^0\pi^0,\, \sqrt{s_V}\, ) \sim
\left|\frac{1}{s_V-M^2_\phi+i\sqrt{s_V}\,
\Gamma_\phi(\sqrt{s_V}\,)}\right|^2 \nonumber \\ && \left|
\frac{g_\pi}{ s_S-M^2_0+ig_\pi^2 \rho_{\pi\pi}(s_S) +ig_K^2\rho_{K\bar
K}(s_S)}+B_{\phi}(m^2_\phi,s_S,0)\right|^2\ .
\label{acha2}
\end{eqnarray}
Hereafter, to avoid possible confusion, we use the notation of our
paper,
 replacing the notations of \cite{acha} as follows: $E^2\to s_V$ ,
 $M_{\pi\pi}^2\to s_S$ and $B(M_{\pi\pi}^2)\to (-)
 B_{\phi}(m^2_\phi,s_S,0)$.

The formula (\ref{acha2}) is incorrect. Correct formula for the
amplitude $e^+e^-\to \gamma\pi^0\pi^0$ is given by eq. (\ref{19a})
which is characterised by four terms, with corresponding substitution
of the pole factors:
\begin{eqnarray} &&(s_V- m^2_\phi)^{-1} \to
\left(s_V-M^2_\phi+iM_\phi\Gamma_\phi\right)^{-1}\, ,
\nonumber \\
&&
(s_V- m^2_\phi)^{-1} \to
 \left(
s_S-M^2_0+ig_\pi^2 \rho_{\pi\pi}(s_S)
+ig_K^2\rho_{K\bar K}(s_S)\right)^{-1} .
\label{acha3-1}
\end{eqnarray}
The threshold theorem is given by (\ref{19a-t}), it does not lead to
eq. (10) of Achasov's paper \cite{acha}.

 \section{Conclusion}
In Section 1, I present the logic of calculation of the reactions
$e^+e^-\to  \gamma\pi\pi$ ,
$ \phi \to \gamma f_0$ and $ \phi \to \gamma\pi\pi $ which
was accepted
in \cite{1} (and in  previous papers \cite{epja,s}). Also, the main
formulae for the considered reactions are given for the case of a
description
of resonances by the Breit-Wigner poles as well as in the K-matrix
technique.

The formulae presented in Section 1  have nothing common with
those N.N. Achasov claims to be  ours. His criticism is therefore
misplaced.

I thank S.F. Tuan for bringing my attention to Achasov's paper.

\end{document}